\documentclass[sigconf]{acmart}
\AtBeginDocument{%
  }

\copyrightyear{2026}
\acmYear{2026}
\setcopyright{cc}
\setcctype{by}
\acmConference[WCCCE 2026]{Western Canada Conference on Computing Education}{April 30-May 01, 2026}{Vancouver, BC, Canada}
\acmBooktitle{Western Canada Conference on Computing Education (WCCCE 2026), April 30-May 01, 2026, Vancouver, BC, Canada}
\acmDOI{10.1145/3828792.3828814}
\acmISBN{979-8-4007-2755-9/2026/04}

\begin{document}

\title{Looking Back and Forward: What Teaching Materials Do Not Remember About Instructional Reasoning}

\author{Yi Ching Chou}
\email{ycchou@sfu.ca}
\affiliation{%
  \institution{Simon Fraser University}
  \city{Burnaby}
  \state{British Columbia}
  \country{Canada}
}

\author{Hao Fang}
\email{fanghaof@sfu.ca}
\affiliation{%
  \institution{Simon Fraser University}
  \city{Burnaby}
  \state{British Columbia}
  \country{Canada}
}

\author{Haoyuan Zhao}
\email{hza127@sfu.ca}
\affiliation{%
  \institution{Simon Fraser University}
  \city{Burnaby}
  \state{British Columbia}
  \country{Canada}
}

\author{Jiangchuan Liu}
\email{jcliu@sfu.ca}
\affiliation{%
  \institution{Simon Fraser University}
  \city{Burnaby}
  \state{British Columbia}
  \country{Canada}
}



\renewcommand{\shortauthors}{Chou et al.}

    

\begin{abstract}
Instructional materials can be passed along to new instructors, yet the instructional reasoning, such as instructional intentions, contextual information, and reflective insights behind their design, is rarely preserved. Traditional Learning Management System (LMS) tools are designed to store instructional artifacts rather than reasoning. To address this issue, we propose Teaching Memory, a design approach that treats instructional reasoning as a primary form of knowledge. We present a design-oriented framework for preserving instructional reasoning in a lightweight manner within instructors’ existing workflows, enabling continuity and reuse.
\end{abstract}

\ccsdesc[500]{Applied computing~Learning management systems}

\keywords{Preserving Instructional Reasoning, Computer Science Education, Teaching Memory}

\maketitle

\section{Introduction}

Well-designed teaching materials are often shared among instructors and reused across course offerings. However, these materials tend to persist only as instructional artifacts. While the materials themselves can be easily passed along, the instructional reasoning, such as instructional intentions, contextual information, and reflective insights, is rarely well-documented and often lost when the creators leave the institution. New instructors will need to infer the instructional reasoning, leading to misalignment with the original pedagogical focus or unnecessary onboarding effort.

Existing Learning Management System (LMS) tools are primarily designed to store the outcomes of instructional design, such as lecture slides and assignments. These outcomes rarely capture the instructional reasoning since the instructional reasoning is often developed incrementally and is highly context-dependent, affected by class size, curriculum constraints, student background, and the teaching philosophy of the creators. Unfortunately, this information is typically not explicitly stored in LMS tools.

This gap suggests the need for a new design approach beyond traditional LMS tools to preserve instructional reasoning at the institutional level. In this paper, we refer to the new approach as Teaching Memory, which treats instructional reasoning as a primary form of knowledge. Teaching Memory aims to capture the instructional intentions, contextual information, and reflective insights embedded in instructional reasoning, enabling long-term continuity and knowledge transfer within the institution.

\section{A Design Proposal for Teaching Memory}

Teaching Memory is designed as a new approach to preserve instructional reasoning. In contrast to traditional Learning Management System (LMS) tools, which primarily store teaching materials as the outcomes of instructional design, Teaching Memory focuses on the instructional reasoning that drives instructional design decisions. Teaching Memory focuses on several conceptual components that preserve instructional reasoning.

\textbf{Instructional intentions.} Instructional intentions guide instructional design decisions, such as why the instructor emphasizes certain concepts and chooses to present the materials in certain ways. These intentions are often implicit in the teaching materials. By preserving explicit instructional intentions in Teaching Memory, future instructors can understand the pedagogical focus behind existing teaching materials \cite{Stefaniak2021-bk} rather than inferring intentions. This enables new instructors to adapt the existing materials while maintaining the pedagogical focus.

\textbf{Contextual information.} Contextual information describes the conditions and constraints that the instructor was facing when designing the teaching materials, such as class size, student background, and deadline. The contextual information significantly influences how teaching materials are designed, created, and presented, yet it is rarely explicitly documented. Teaching Memory preserves contextual information, enabling new instructors to adapt teaching materials under different contexts.

\textbf{Reflective
insights.} Reflective insights capture the lessons learned and observations under specific instructional contexts, such as which topics students commonly struggled with and which explanations were effective. Preserving these reflective insights in Teaching Memory can enable instructors to revisit prior teaching experience and refine their current instructional intentions.

Teaching Memory consolidates the instructional intentions, contextual information, and reflective insights to preserve instructional reasoning. This allows the institution to maintain long-term continuity of teaching knowledge across instructors, course offerings, and semesters.

\section{Teaching Memory From Past to Future}

\begin{figure}[t]
    \centering
    \includegraphics[width=0.6\linewidth]{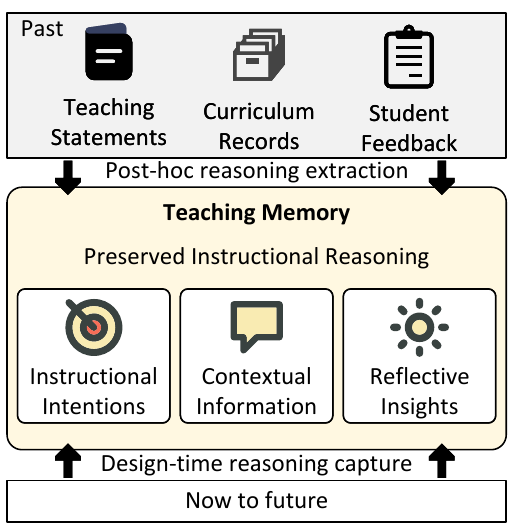}
    \caption{Teaching memory from past to future.}
    \label{fig:teaching_memory}
    \vspace{-3mm}
\end{figure}

Teaching Memory is proposed as a design pattern rather than a specific tool implementation. This section discusses how Teaching Memory can be constructed across time, from existing instructional materials to future teaching practices. 

As illustrated in Figure~\ref{fig:teaching_memory}, instructional reasoning can be reconstructed from past materials, such as teaching statements, curriculum records, and student feedback. Moving forward, Teaching Memory enables instructional reasoning to be captured at design time as instructors make instructional decisions.

\subsection{Teaching Memory From the Past} 
Many institutions have already accumulated a large amount of valuable teaching materials. These materials must be analyzed to elicit their instructional reasoning. Three questions must be answered for effective instructional reasoning elicitation:

\begin{enumerate}
    \item Who is (are) the creator(s) of the teaching materials?
    \item What is their pedagogical philosophy and focus?
    \item What is the contextual information and reflective insights of designing those teaching materials?
\end{enumerate}

The answers to (1) and (2) are embedded in \emph{teaching statements}, which most of the teaching faculty have their teaching statement available. In teaching statements, instructors often state their teaching philosophy and teaching approach while some provide further explanation in the statements. These statements can be good sources for extracting the instructional intentions of the instructors. While not all past contextual information and reflective insights are available, some of them can be extracted from curriculum records and student feedback to answer (3).

\subsection{Teaching Memory From Now to Future} 
For Teaching Memory to be sustainable, it must motivate instructors to preserve instructional reasoning while imposing minimal additional burden. To achieve this, the design of Teaching Memory for future teaching materials should follow three principles:

\begin{enumerate}
    \item The capture of instructional reasoning should be during instructional decision-making.
    \item The capture of instructional reasoning should be low friction and embedded within existing instructional practices.
    \item Instructional reasoning should be revisited for continuous review and refinement.
\end{enumerate}

Preserving instructional reasoning can be naturally integrated into instructors' existing workflows during instructional decision-making while maintaining low friction. For example, when instructors are designing lecture slides, they can record brief notes in the speaker notes to explain the instructional intentions of a slide, the contextual information about the class and course, and the reflective insights about the old version of the slides being used in the class. These lightweight notes capture instructional reasoning at the moment of design.

In addition, as the use of generative AI tools to create teaching materials becomes more common \cite{ALIER2025103940}, instructional reasoning can also be captured through conversational interactions with AI tools. At the end of a conversation, a reflective prompt \cite{ALGHAMDI2026105511} can be used to summarize the instructional intentions, contextual information, and reflective insights underlying the design choices. This summary can then be reused in future conversations, enabling instructors to revisit and refine their instructional reasoning over time.

\section{Discussion and Future Work}
Teaching Memory is presented as a design proposal in this paper rather than a fully implemented system. Teaching Memory focuses on lightweight preservation of instructional reasoning at the moment of designing teaching materials rather than documenting everything to mitigate additional burden on instructors.

Future work includes exploring possible implementations that integrate Teaching Memory into existing instructional tools and studying how instructors adopt and interact with Teaching Memory in practice. Evaluating the long-term impact of Teaching Memory on students, instructors, and institutions remains an important direction for our future work. 



\bibliographystyle{ACM-Reference-Format}
\bibliography{references}

@ARTICLE{Stefaniak2021-bk,
  title    = "Fostering pedagogical reasoning and dynamic decision-making
              practices: a conceptual framework to support learning design in a
              digital age",
  author   = "Stefaniak, Jill and Luo, Tian and Xu, Meimei",
  journal  = "Educational Technology Research and Development",
  volume   =  69,
  number   =  4,
  pages    = "2225--2241",
  month    =  aug,
  year     =  2021
}

@article{ALIER2025103940,
title = {LAMB: An open-source software framework to create artificial intelligence assistants deployed and integrated into learning management systems},
journal = {Computer Standards \& Interfaces},
volume = {92},
pages = {103940},
year = {2025},
issn = {0920-5489},
doi = {https://doi.org/10.1016/j.csi.2024.103940},
url = {https://www.sciencedirect.com/science/article/pii/S0920548924001090},
author = {Marc Alier and Juanan Pereira and Francisco José García-Peñalvo and Maria Jose Casañ and Jose Cabré}
}

@article{ALGHAMDI2026105511,
title = {Leveraging prompt-based LLMs for automated scoring and feedback generation in higher education},
journal = {Computers \& Education},
volume = {243},
pages = {105511},
year = {2026},
issn = {0360-1315},
doi = {https://doi.org/10.1016/j.compedu.2025.105511},
url = {https://www.sciencedirect.com/science/article/pii/S0360131525002799},
author = {Eman Mudhi AlGhamdi and Yuheng Li and Dragan Gašević and Guanliang Chen}
}

@String{Computer = "{IEEE} Computer" }

\end{document}